\documentclass[aps,preprint]{revtex4}%
\usepackage[latin9]{inputenc}
\usepackage{amsmath}
\usepackage{amssymb}
\usepackage{amsfonts}
\usepackage{graphicx}%
\usepackage{natbib}
\setcitestyle{super}

\begin{document}
\title{Local pressure for inhomogeneous fluids}
\author{James Dufty}
\email{dufty@ufl.edu}
\affiliation{Department of Physics, University of Florida, Gainesville, FL
32611, USA}
\author{Jeffrey Wrighton}
\email{wrighton@ufl.edu} 
\affiliation{Department of Physics, University of Florida, Gainesville, FL
32611, USA}
\author{Kai Luo}
\email{kluo@carnegiescience.edu}
\affiliation{Earth and Planets Laboratory, Carnegie Institution for Science,
Washington, DC 20015-1305, USA}

\begin{abstract}
Definitions for a local pressure in an inhomogeneous fluid are considered for
both equilibrium and local equilibrium states. Thermodynamic and mechanical
(hydrodynamic) contexts are reconciled. Remaining problems and uncertainties
are discussed.

\end{abstract}
\date{\today}
\maketitle

\section{Introduction}

\label{sec1}The global pressure for equilibrium systems is well-defined within
statistical mechanics, both quantum and classical \cite{Huang}. For extensive
systems the definitions for different ensembles are equivalent. Here the grand
canonical ensemble will be chosen for such representations, characterized by
an inverse temperature $\beta$ and activity $\nu$. The global pressure is
then proportional to the grand potential which defines all thermodynamic
properties of the system. Its form is obtained from variations of the grand
potential with respect to volume, leading to the equilibrium average of a
specific operator equivalent to the familiar virial equation. For the special
case of inhomogeneous equilibrium states for systems with an external
potential $v^{{\mathrm{ext}}}\left(  \mathbf{r}\right)  $ the effective
activity $\nu\left(  \mathbf{r}\right)  \equiv\nu-\beta v^{{\mathrm{ext}}}\left(
\mathbf{r}\right)  $ varies locally so it is appropriate to define an
associated local thermodynamics \cite{Evans92}. The first objective here is to
explore how to define a \textit{local }thermodynamic pressure $p^{e}\left(
\mathbf{r},\beta\mid\nu\right)  $ whose spatial integral is the global
pressure \cite{Percus82,Gubbins93,Percus86}. The notation indicates that the
local pressure depends on the space point $\mathbf{r}$, the inverse
temperature $\beta$, and is a functional of $\nu\left(  \mathbf{r}\right)  $.
It is straightforward to identify a local operator whose ensemble average
integrates to the global virial equation, defining such a local pressure.
However, it is not unique since any contribution whose integral vanishes can
be added. Additional constraints are needed.

A conceptually different notion of pressure is obtained from the average
momentum flux of the inhomogeneous equilibrium fluid, or pressure tensor
$p_{ij}^{e}\left(  \mathbf{r},\beta\mid\nu\right)  $. Local conservation of
momentum at equilibrium leads to a force balance equation relating the
gradient of that pressure tensor to the applied external force. This approach
has an extensive history in the classical description of interfaces and
surface tension \cite{Irving50,Harasima58,Schofield82,Walton83}. Derivation of 
the conservation law from
the underlying Heisenberg dynamics provides the form of the operator whose
average gives the pressure tensor. It does not explicitly exploit the grand
potential or any thermodynamics other than the stationarity of the equilibrium
state. While the local thermodynamic pressure is defined only indirectly from
the global pressure, the pressure tensor is inherently a local property.
However, this method only provides the divergence of the pressure tensor and
the latter is therefore not unique. Consequently the related scalar pressure
$p_{m}^{e}\left(  \mathbf{r},\beta\mid\nu\right)  \equiv p_{ii}^{e}\left(
\mathbf{r},\beta\mid\nu\right)  /3$ also is not unique (here and below a
summation over repeated indices is implied). The pressure obtained from the
pressure tensor will be referred to as the hydrodynamic or mechanical pressure
as it appears in the macroscopic conservation equations. Clearly, it is
desirable that the thermodynamic and hydrodynamic pressures should be the same
for consistent representations of the stationary states. It is expected that
the uncertainties in each can be exploited to assure this equivalence.

Two cases are considered here, the inhomogeneous equilibrium states described
above, and their generalization to local equilibrium states. The latter differ
in the sense that the inverse temperature can be spatially varying,
$\beta=\beta\left(  \mathbf{r}\right)$, in addition to the activity
$\nu\left(  \mathbf{r}\right)  $. For equilibrium states it is shown that the
uncertainty in the thermodynamic pressure can be removed by adding a
contribution that equates it to the hydrodynamic pressure.

The same objective arises in the more general context of non-stationary
\textit{local equilibrium} states of hydrodynamics. The associated ensemble is
similar to the equilibrium ensemble. A ``thermodynamics" for this state can be
defined from the associated grand potential and an associated local pressure
identified \cite{Dufty20,Zubarev}. However, in this case, for spatially
varying $\beta\left(  \mathbf{r}\right)  $, there is no longer the flexibility
to modify the thermodynamic local pressure to be equal to that from the local
equilibrium average of the stress tensor. Consequently, it would seem that the
equation of state for hydrodynamics is not the same as that for local
equilibrium thermodynamics. The precise difference is identified below.
Unfortunately, this implies that the equivalence chosen for the strict
equilibrium noted above is not recovered from the hydrodynamic equations for
their stationary\ limit.

This paradox is resolved by exploiting the uncertainty in the pressure tensor.
A divergenceless additional contribution to the average momentum flux can be
chosen such that the local pressure associated with the new momentum flux
agrees with the thermodynamic pressure. In this way thermodynamic and
mechanical concepts are reconciled.

The analysis here is based in quantum statistical mechanics so that all
average properties have an associated underlying operator representing them.
The calculation of their averages is not discussed but the connection to
density functional theory methods is indicated. An alternative approach is to
postulate an average pressure tensor and verify that it yields the required
macroscopic force balance equation. This has been described by Percus
\cite{Percus86} for the inhomogeneous equilibrium fluid. His pressure tensor
is entirely characterized by the thermodynamic free energy density. It is
described in Supporting Information S.3 and the associated pressure is identified.

The primary importance of this investigation of equivalence is for the local equilibrium states
of hydrodynamics. In that case both concepts of the pressure occur. The first is as a functional
relationship between the fundamental conserved number, energy, and momentum densities and
their conjugate fields activity, temperature, and flow velocity. This functional relationship is the
thermodynamics of the local equilibrium grand potential, or thermodynamics pressure. The second
occurrence is through the average momentum flux, comprised as a reference local equilibrium
average and a dissipative component. Only the reference contribution is considered here and is
referred to as the average mechanical pressure tensor here. It is a functional of the conjugate fields.
Hence the equivalence of the thermodynamic and mechanical pressures is a necessary condition for
the hydrodynamic equations to provide a closed local macroscopic description, regardless of the
choice for the dissipative component (e.g., Navier-Stokes or far from equilibrium).

At this point it is appropriate to characterize the context by noting related topics not
bearing directly on the question of equivalence. The lack of uniqueness for the mechanical pressure
tensor is well-known; for early references see Refs. \citenum{Irving50,Harasima58,Schofield82,Walton83}. Most of these studies do not make explicit the
required equivalence with thermodynamic pressure. An exception is one demonstration that the
Harasima choice gives the wrong pressure in cylindrical coordinates\cite{Hafskjold02}. This is resolved in a recent
work for cylindrical geometry by synthesizing the Irving and Kirkwood and Harasima expressions
for different coordinates\cite{Shi20}. Another definition of the local thermodynamic pressure closer to that
given here\cite{Lovett97} does not make any connection to the various choices for the mechanical pressure
tensor. In summary, the work here is complementary to this important body of work by relating the
two different studies of thermodynamic pressure and mechanical pressure tensor. Further comment
is given in the final section where measurement by simulation is briefly discussed.

The local equilibrium thermodynamics considered here describes a reference state for real
non-equilibrium systems. In the context of information entropy\cite{Jaynes57a,Jaynes57b,Robertson} it provides the optimal
representation in terms of the given exact average local conserved densities. The latter must be
provided from some detailed exact theory (e.g., Liouville- von Neumann equation). The grand
potential associated with the local equilibrium ensemble provides the ``equation of state" for
generating conjugate variables like temperature and activity, and is the direct analogue of equilibrium
thermodynamics. To further clarify the context it is noted that the current work does not relate to
the general fields of ``non-equilibrium thermodynamics"\cite{Van20} nor ``extended thermodynamics"\cite{Jou20}.
The former is an attempt to discover universal fundamental principles, similar to those of
equilibrium thermodynamics (e.g., a generalized second law, entropy), to govern the dissipative
dynamics of macroscopic properties. Extended irreversible thermodynamics takes as the
macroscopic fields the usual local conserved fields plus the dissipative fluxes of energy and
momentum. The conservation laws of ordinary hydrodynamics must then be supplemented with
unknown additional equations for the dissipative fluxes. If the latter could be given in terms of the
conserved fields this would not be necessary, so in a sense extended hydrodynamics is a tool to
discover those forms. The local equilibrium thermodynamics considered here is not a theory, such as
those sought above, but rather an exact functional relationship among two equivalence classes of
fields --- there is no entropy production nor inherent dissipation beyond that of the input fields.
While it represents general non-equilibrium states, it is not predictive without the hydrodynamic
equations themselves (for an exact formulation of the latter see Ref. \citenum{Dufty20}).

\section{Local pressure for an inhomogeneous fluid at equilibrium}

\label{sec2}Consider first a system of $N$ particles in a large volume $V$
with Hamiltonian%
\begin{equation}
\mathcal{H}_{N}=H_{N}+\sum_{\alpha=1}^{N}v^{{\mathrm{ext}}}({\mathbf{q}%
}_{\alpha})), \label{2.1}%
\end{equation}
where $v^{{\mathrm{ext}}}({\mathbf{q}}_{\alpha})$ is an external potential
coupling to the particle with position operator ${\mathbf{q}}_{\alpha}$, and
the isolated system Hamiltonian $H_{N}$ is
\begin{equation}
H_{N}=\sum_{\alpha=1}^{N}\frac{p_{\alpha}^{2}}{2m}+\frac{1}{2}\sum_{\alpha
\neq\sigma=1}^{N}U_{N}(\left\vert {\mathbf{q}}_{\alpha}-{\mathbf{q}}_{\sigma
}\right\vert ). \label{2.2}%
\end{equation}
Here $U_{N}(\left\vert {\mathbf{q}}_{\alpha}-{\mathbf{q}}_{\sigma}\right\vert
)$ is a pair potential for particles $\alpha$ and $\sigma$, and ${\mathbf{p}%
}_{\alpha}$ is the momentum operator for particle $\alpha$. At equilibrium
with inverse temperature $\beta$ and activity $\nu$ the average of a property
characterized by an operator $X_{N}$ is given in the grand canonical ensemble
by
\begin{equation}
\left\langle X\right\rangle ^{e}\equiv\sum_{N}Tr^{(N)}X_{N}\rho_{N}%
^{e},\hspace{0.25in}\rho_{N}^{e}=e^{-Q^{e}\left(  \beta,V\mid\nu\right)
}e^{-\left(  \beta H_{N}-\int d\mathbf{r}\nu\left(  \mathbf{r}\right)
n\left(  \mathbf{r}\right)  \right)  }. \label{2.3}%
\end{equation}
The external potential has been combined with the activity to define a local
activity $\nu\left(  \mathbf{r}\right)  $%
\begin{equation}
\nu\left(  \mathbf{r}\right)  \equiv\nu\mathbf{-}\beta v^{{\mathrm{ext}}}\left(
\mathbf{r}\right)  , \label{2.4}%
\end{equation}
and $n\left(  \mathbf{r}\right)  $ is the number density operator%
\begin{equation}
n\left(  \mathbf{r}\right)  =\sum_{\alpha=1}^{N}\delta\left(  \mathbf{r}%
-\mathbf{q}_{\alpha}\right)  . \label{2.5}%
\end{equation}
The notation $Y\left(  \beta,V\mid\nu\right)  $ here and below denotes a
function of $\beta,V$ and a functional of $\nu\left(  \mathbf{r}\right)  .$
The normalization function $Q^{e}\left(  \beta,V\mid\nu\right)  $ is known as
the grand potential \cite{terHaar61} and is chosen such that $\left\langle
1\right\rangle ^{e}=1$%
\begin{equation}
Q^{e}\left(  \beta,V\mid\nu\right)  =\ln\sum_{N}Tr^{(N)}e^{-\left(  \beta
H_{N}-\int d\mathbf{r}\nu\left(  \mathbf{r}\right)  n\left(  \mathbf{r}%
\right)  \right)  }\,. \label{2.6}%
\end{equation}
It determines the complete thermodynamics for the system through the
definition of the global pressure
\begin{equation}
\beta P^{e}\left(  \beta\mid\nu\right)  V=Q^{e}\left(  \beta,V\mid\nu\right)
. \label{2.7}%
\end{equation}
$.$

For a sufficiently large volume $Q^{e}\left(  \beta,V\mid\nu\right)  $ is
extensive (proportional to $V$) so the pressure is independent of the volume.
Then an equivalent form for (\ref{2.7}) is%
\begin{equation}
\beta P^{e}\left(  \beta\mid\nu\right)  =\left.\frac{\partial Q^{e}\left(
\beta,V\mid\nu\right)  }{\partial V}\right|_{\beta,\nu}. \label{2.8}%
\end{equation}
The volume derivative can be calculated directly (e.g., using length scaling
\cite{Dufty16,Dufty20}) to get%
\begin{equation}
P^{e}\left(  \beta\mid\nu\right)  =\frac{1}{3V}\left(  2\left\langle
K\right\rangle ^{e}+\left\langle \mathcal{V}\right\rangle ^{e}\right)  ,
\label{2.9}%
\end{equation}
where $K$ is the kinetic energy operator and $\mathcal{V}$ is the virial
operator (for the internal forces)%
\begin{equation}
K=\sum_{\alpha=1}^{N}\frac{1}{2m}p_{\alpha j}^{2},\hspace{0.25in}%
\mathcal{V=}\frac{1}{2}\sum_{\alpha\neq\gamma=1}^{N}({\mathbf{q}}_{\gamma
}-{\mathbf{q}}_{\alpha})\cdot\mathbf{F}_{\alpha\gamma}(\left\vert {\mathbf{q}%
}_{\alpha}-{\mathbf{q}}_{\gamma}\right\vert ). \label{2.10}%
\end{equation}
It is seen that (\ref{2.9}) is the usual virial equation for the global
(intrinsic) pressure, confirming the consistency of the thermodynamic and
mechanical concepts of global pressure.

The objective now is to identify an associated local density pressure. It is
done by defining a local density for the grand potential in (\ref{2.7})
\begin{equation}
\int d\mathbf{r}\beta p^{e}\left(  \mathbf{r},\beta\mid\nu\right)
=Q^{e}\left(  \beta,V\mid\nu\right)  . \label{2.10d}%
\end{equation}
Accordingly, $V^{-1}p^{e}\left(  \mathbf{r},\beta\mid\nu\right)  $ is the
local density for the global pressure
\begin{equation}
P^{e}\left(  \beta\mid\nu\right)  \equiv\frac{1}{V}\int d\mathbf{r}%
p^{e}\left(  \mathbf{r},\beta\mid\nu\right)  . \label{2.11}%
\end{equation}
One choice to identify it is to replace the operators $K$ and $\mathcal{V}$
\ in (\ref{2.9}) by associated local densities%
\begin{equation}
p_{0}^{e}\left(  \mathbf{r},\beta\mid\nu\right)  =\frac{1}{3}\left\langle
2K_{0}({\mathbf{r}})+\mathcal{V}_{0}({\mathbf{r}})\right\rangle ^{e},
\label{2.10a}%
\end{equation}
where%
\begin{equation}
K_{0}({\mathbf{r}})=\frac{1}{4m}\sum_{\alpha=1}^{N}\left[  p_{\alpha}%
^{2},\delta\left(  \mathbf{r-q}_{\alpha}\right)  \right]  _{+} \label{2.10b}%
\end{equation}%
\begin{equation}
\mathcal{V}_{0}({\mathbf{r}})=\frac{1}{2}\sum_{\alpha\neq\sigma=1}%
^{N}F_{\alpha\sigma i}\left(  \left\vert \mathbf{q}_{\alpha}-\mathbf{q}%
_{\sigma}\right\vert \right)  ({\mathbf{q}}_{\sigma}-{\mathbf{q}}_{\alpha
})\delta\left(  \mathbf{r-q}_{\alpha}\right)  . \label{2.10c}%
\end{equation}
The brackets $[a,b]_{+}$ denote an anti-commutator. It is required in
(\ref{2.10b}) to assure that $K_{0}({\mathbf{r}})$ is Hermitian. More
generally, the local pressure can be expressed as%
\begin{equation}
p^{e}\left(  \mathbf{r},\beta\mid\nu\right)  =p_{0}^{e}\left(  \mathbf{r}%
,\beta\mid\nu\right)  +\Delta p_{0}^{e}\left(  \mathbf{r},\beta\mid\nu\right)
\label{2.11a}%
\end{equation}
where $\Delta p_{0}^{e}\left(  \mathbf{r},\beta\mid\nu\right)  $ is any
functional whose volume integral vanishes.
\begin{equation}
\int d\mathbf{r}\Delta p^{e}\left(  \mathbf{r},\beta\mid\nu\right)  =0.
\label{2.11b}%
\end{equation}

To suggest an alternative choice for $p^{e}\left(  \mathbf{r},\beta\mid
\nu\right)  $ consider the exact microscopic conservation law for the momentum
density operator $p_{i}({\mathbf{r}},t)$ (see Supporting Information S.1)%
\begin{equation}
\partial_{t}p_{i}({\mathbf{r}},t)+\partial_{j}t_{ij}({\mathbf{r}%
},t)=-n({\mathbf{r}},t)\partial_{i}v^{{\mathrm{ext}}}({\mathbf{r}},t)\,,
\label{2.12}%
\end{equation}
where the momentum density operator is%
\begin{equation}
\mathbf{p}(\mathbf{r},t)=\frac{1}{2}\sum_{\alpha=1}^{N}\left[  \mathbf{p}%
_{\alpha}\left(  t\right)  ,\delta(\mathbf{r}-\mathbf{q}_{\alpha}\left(
t\right)  )\right]  _{+} \label{2.13}%
\end{equation}
and the momentum flux is%
\begin{align}
t_{ij}\left(  {\mathbf{r,}}t\right)   &  =\frac{1}{4m}\sum_{\alpha=1}%
^{N}\left[  p_{i\alpha}\left(  t\right)  ,\left[  p_{j\alpha}\left(  t\right)
,\delta\left(  \mathbf{r-q}_{\alpha}\left(  t\right)  \right)  \right]
_{+}\right]  _{+}\nonumber\\
&  +\frac{1}{2}\sum_{\alpha\neq\sigma=1}^{N}F_{\alpha\sigma i}\left(
\left\vert \mathbf{q}_{\alpha}\left(  t\right)  -\mathbf{q}_{\sigma}\left(
t\right)  \right\vert \right)  \mathcal{D}_{j}\left(  \mathbf{r,q}_{\alpha
}\left(  t\right)  ,{\mathbf{q}}_{\sigma}\left(  t\right)  \right)  .
\label{2.14}%
\end{align}
The operator $\mathcal{D}_{j}\left(  \mathbf{r,q}_{\alpha}\left(  t\right)
,{\mathbf{q}}_{\sigma}\left(  t\right)  \right)  $ is given by%
\begin{equation}
\mathcal{D}_{j}\left(  \mathbf{r,q}_{1},\mathbf{q}_{2}\right)  \equiv
\int_{\mathcal{C}}d\lambda\frac{dx_{j}\left(  \lambda\right)  }{d\lambda
}\delta\left(  \mathbf{r-x}\left(  \lambda\right)  \right)  ,\hspace
{0.26in}\mathbf{x}\left(  \lambda_{1}\right)  =\mathbf{q}_{1},\hspace
{0.26in}\mathbf{x}\left(  \lambda_{2}\right)  =\mathbf{q}_{2}. \label{2.15}%
\end{equation}
Here $\mathcal{C}$ is an arbitrary continuous path connecting $\mathbf{x}%
\left(  \lambda\right)  $ between $\lambda_{1}$ and $\lambda_{2}.$ (Further
comment on the choice is given in Supporting Information S.2.)

Since the equilibrium ensemble is stationary, the equilibrium average of
(\ref{2.12}) gives the stationary equilibrium force balance equation%
\begin{equation}
\partial_{j}\left\langle t_{ij}({\mathbf{r}})\right\rangle ^{e}=-\left\langle
n({\mathbf{r}})\right\rangle ^{e}\partial_{i}v^{{\mathrm{ext}}}({\mathbf{r}})
\label{2.16}%
\end{equation}
This is the expected local stability condition. A mechanical pressure can be
associated with the average momentum flux according to%
\begin{equation}
p_{m}^{e}\left(  \mathbf{r},\beta\mid\nu\right)  \equiv\frac{1}{3}\left\langle
t_{ii}({\mathbf{r}})\right\rangle ^{e} \label{2.16a}%
\end{equation}
Note that (\ref{2.14}) implies
\begin{equation}
\frac{1}{V}\int d{\mathbf{r}}\frac{1}{3}\left\langle t_{ii}({\mathbf{r}%
})\right\rangle ^{e}=\frac{1}{3V}\left(  2\left\langle K\right\rangle
^{e}+\left\langle \mathcal{V}\right\rangle ^{e}\right)  =P^{e}\left(
\beta\mid\nu\right)  \label{2.17}%
\end{equation}
where the last equality follows from (\ref{2.9}). Use has been made of the
identity
\begin{equation}
\int d{\mathbf{r}}\mathcal{D}({\mathbf{r}},{\mathbf{q}}_{\alpha},{\mathbf{q}%
}_{\gamma})={\mathbf{q}}_{\alpha}-{\mathbf{q}}_{\gamma}. \label{2.18}%
\end{equation}
The left side of (\ref{2.17}) suggests an alternative choice for the
definition of a local pressure in (\ref{2.11})
\begin{equation}
p^{e}\left(  \mathbf{r},\beta\mid\nu\right)  =p_{m}^{e}\left(  \mathbf{r}%
,\beta\mid\nu\right)  =\frac{1}{3}\left\langle t_{ii}({\mathbf{r}%
})\right\rangle ^{e}. \label{2.19}%
\end{equation}
This definition has a mechanical origin, without direct reference to the grand
potential or thermodynamics.

The two choices $p_{0}^{e}\left(  \mathbf{r},\beta\mid\nu\right)  $ and
$p_{m}^{e}\left(  \mathbf{r},\beta\mid\nu\right)  $ are clearly different, but
both yield the thermodynamic global pressure. The choice $p^{e}\left(
\mathbf{r},\beta\mid\nu\right)  =p_{m}^{e}\left(  \mathbf{r},\beta\mid
\nu\right)  $ for the local pressure in (\ref{2.11}) is clearly the desirable
one as it assures the mechanical balance equation is consistent with thermodynamics.

Since $t_{ii}({\mathbf{r}})$ has a precise microscopic origin, this provides a
local microscopic basis for the pressure%

\begin{equation}
p^{e}\left(  \mathbf{r},\beta\mid\nu\right)  =\frac{1}{3}\left(  2\left\langle
K({\mathbf{r}})\right\rangle ^{e}+\left\langle \mathcal{V}({\mathbf{r}%
})\right\rangle ^{e}\right)  \label{2.20d}%
\end{equation}
with%
\begin{equation}
K({\mathbf{r}})=\frac{1}{8m}\sum_{\alpha=1}^{N}\left[  p_{i\alpha},\left[
p_{i\alpha},\delta\left(  \mathbf{r-q}_{\alpha}\right)  \right]  _{+}\right]
_{+} \label{2.20e}%
\end{equation}
and
\[
\mathcal{V}({\mathbf{r}})=\frac{1}{2}\sum_{\alpha\neq\sigma=1}^{N}%
F_{\alpha\sigma i}\left(  \left\vert \mathbf{q}_{\alpha}-\mathbf{q}_{\sigma
}\right\vert \right)  \mathcal{D}_{i}\left(  \mathbf{r,q}_{\alpha}%
,{\mathbf{q}}_{\sigma}\right)
\,.
\]
This is the desired result.

It is seen that while the global pressure is the same for $p^{e}\left(
\mathbf{r},\beta\mid\nu\right)  $ and $p_{0}^{e}\left(  \mathbf{r},\beta
\mid\nu\right)  $, the local pressures differ in their microscopic realizations%

\begin{equation}
K({\mathbf{r}})=K_{0}\left(  \mathbf{r}\right)  +\frac{\hbar^{2}}{8m}%
\nabla^{2}n\left(  \mathbf{r}\right)  \,, \label{2.24}%
\end{equation}%
\begin{equation}
\mathcal{V}(\mathbf{r})=\mathcal{V}_{0}({\mathbf{r}})+\frac{1}{2}\sum
_{\alpha\neq\sigma=1}^{N}F_{\alpha\sigma i}\left(  \left\vert \mathbf{q}%
_{\alpha}-\mathbf{q}_{\sigma}\right\vert \right)  \left[  \mathcal{D}%
_{i}\left(  \mathbf{r,q}_{\alpha},{\mathbf{q}}_{\sigma}\right)  -({\mathbf{q}%
}_{\sigma}-{\mathbf{q}}_{\alpha})_{i}\delta\left(  \mathbf{r-q}_{\alpha
}\right)  \right] \,. \label{2.25}%
\end{equation}
To get (\ref{2.24}) use has been made of the identity%
\begin{equation}
\frac{1}{2m}\sum_{\alpha}p_{\alpha i}\delta\left(  \mathbf{r}-\mathbf{q}%
_{\alpha}\right)  p_{\alpha i}=K_{0}\left(  \mathbf{r}\right)  +\frac
{\hbar^{2}}{4m}\nabla^{2}n\left(  \mathbf{r}\right)  . \label{2.26}%
\end{equation}
The dependence on $\hbar$ has been made explicit in (\ref{2.24}) and
(\ref{2.26}) to emphasize this is a purely quantum effect. The identification
of the local pressure in terms of the average momentum flux is a common
definition. What is new here is its identification as the thermodynamic local
pressure for the grand ensemble.

The local pressure $p_{m}^{e}\left(  \mathbf{r},\beta\mid\nu\right)  $ is more
sensitive to spatial variations of the inhomogeneous state than $p_{0}%
^{e}\left(  \mathbf{r},\beta\mid\nu\right)  $. In fact $p_{0}^{e}\left(
\mathbf{r},\beta\mid\nu\right)  $ results from it by a leading order Taylor
series approximation, e.g.%
\begin{align}
\mathcal{D}_{i}\left(  \mathbf{r,q}_{\alpha},{\mathbf{q}}_{\sigma}\right)   &
=({\mathbf{q}}_{\sigma}-{\mathbf{q}}_{\alpha})_{i}\delta\left(  \mathbf{r-q}%
_{\alpha}\right)  +\int_{\lambda_{2}}^{\lambda_{1}}d\lambda\frac{dx_{i}\left(
\lambda\right)  }{d\lambda}\left[  \delta\left(  \mathbf{r-x}\left(
\lambda\right)  \right)  -\delta\left(  \mathbf{r-q}_{\alpha}\right)  \right]
\nonumber\\
&  \simeq({\mathbf{q}}_{\sigma}-{\mathbf{q}}_{\alpha})_{i}\delta\left(
\mathbf{r-q}_{\alpha}\right)  .\label{2.27}%
\end{align}
This gives $\mathcal{V}(\mathbf{r})\simeq\mathcal{V}_{0}({\mathbf{r}})$ and
$p_{m}^{e}\left(  \mathbf{r},\beta\mid\nu\right)  \simeq p_{0}^{e}\left(
\mathbf{r},\beta\mid\nu\right)  $. This leading order approximation is
justified only in the context of states that have smooth spatial variations
over distances of the order of the force range. For extreme conditions, such
as occur for warm, dense matter states it would seem that the form
(\ref{2.14}) must be used for the momentum flux.

In summary, two definitions for a local pressure have been identified,
$p_{0}^{e}\left(  \mathbf{r},\beta\mid\nu\right)  $ and $p_{m}^{e}\left(
\mathbf{r},\beta\mid\nu\right)  $. Each has been identified in terms of the
average of an underlying microscopic operator. They provide the same global
thermodynamics in the sense that both of their volume integrals yield
$P^{e}\left(  \beta\mid\nu\right)  .$ However, at the local level they differ
by $\Delta p_{0}^{e}\left(  \mathbf{r},\beta\mid\nu\right)  $, identified from
(\ref{2.24}) - (\ref{2.26}). The choice of $p^{e}\left(  \mathbf{r},\beta
\mid\nu\right)  =p_{m}^{e}\left(  \mathbf{r},\beta\mid\nu\right)  $ for the
thermodynamic pressure is made on the basis of equating thermodynamic and
mechanical definitions.

So far only the scalar local pressure has been considered. There is no
thermodynamic route to define a local pressure tensor for an inhomogeneous
fluid at equilibrium (see however Supporting Information S.3). Instead it is
identified from (\ref{2.16})%

\begin{equation}
p_{ij}^{e}\left(  \mathbf{r},\beta\mid\nu\right)  \equiv\left\langle
t_{ij}({\mathbf{r}})\right\rangle ^{e}. \label{2.28}%
\end{equation}
By construction it has the form
\begin{equation}
p_{ij}^{e}\left(  \mathbf{r},\beta\mid\nu\right)  =\frac{1}{3}p^{e}\left(
\mathbf{r},\beta\mid\nu\right)  \delta_{ij}+\widetilde{p}_{ij}^{e}\left(
\mathbf{r},\beta\mid\nu\right)  , \label{2.30}%
\end{equation}
(with \ $\widetilde{p}_{ij}^{e}$ is its traceless part) and satisfies the
force balance equation
\begin{equation}
\partial_{j}p_{ij}^{e}\left(  \mathbf{r},\beta\mid\nu\right)  =-\left\langle
n({\mathbf{r}})\right\rangle \partial_{i}v^{{\mathrm{ext}}}({\mathbf{r}}).
\label{2.29}%
\end{equation}
In the next section, attempts to extend the equivalence of thermodynamic and
mechanical concepts to local equilibrium states in the same way lead to
difficulties due to the spatial variations of $\beta({\mathbf{r}},t)$.

\section{Local hydrodynamic pressure}

\label{sec3}Consider now a general non-equilibrium state. The macroscopic
hydrodynamic equations have their origins in averages of the underlying
microscopic conservation laws for \ number density, energy density, and
momentum density, $\left\{  \left\langle n({\mathbf{r}},t)\right\rangle
,\left\langle e({\mathbf{r}},t)\right\rangle ,\left\langle \mathbf{p}%
({\mathbf{r}},t)\right\rangle \right\}  $ \cite{McLennan,Zubarev,Dufty20}. In
particular the hydrodynamic equation resulting from the conservation law for
the momentum density follows from the non-equilibrium average of (\ref{2.12})
\begin{equation}
\partial_{t}\left\langle p_{i}({\mathbf{r}},t)\right\rangle +\partial
_{j}\left\langle t_{ij}({\mathbf{r}},t)\right\rangle =-\left\langle
n({\mathbf{r}},t)\right\rangle \partial_{i}v^{{\mathrm{ext}}}({\mathbf{r}%
},t)\,, \label{3.1}%
\end{equation}
where the brackets now denote a non-equilibrium average%
\begin{equation}
\left\langle X\left(  t\right)  \right\rangle \equiv\sum_{N}Tr^{(N)}X_{N}%
\rho_{N}\left(  t\right)  , \label{3.2}%
\end{equation}
and $\rho_{N}\left(  t\right)  $ is a solution to the Liouville - von Neumann
equation. Traditionally, the momentum density is expressed in terms of a local
flow velocity $\mathbf{u}({\mathbf{r}},t)$ defined by%
\begin{equation}
\left\langle \mathbf{p}({\mathbf{r}},t)\right\rangle \equiv m\left\langle
n({\mathbf{r}},t)\right\rangle \mathbf{u}({\mathbf{r}},t) \label{3.3}%
\end{equation}
and (\ref{3.1}) is written in terms of the momentum flux in the local rest
frame%
\begin{equation}
\left\langle t_{ij}({\mathbf{r}},t)\right\rangle =m\left\langle n({\mathbf{r}%
},t)\right\rangle u_{i}({\mathbf{r}},t)u_{j}({\mathbf{r}},t)+\left\langle
t_{0ij}({\mathbf{r}},t)\right\rangle . \label{3.4}%
\end{equation}
Here the rest frame momentum flux $t_{0ij}({\mathbf{r}},t)$ has the same form
as (\ref{2.14}) with the particle momenta in the rest frame, $p_{i\alpha
}\left(  t\right)  \rightarrow p_{i\alpha}\left(  t\right)  -mu_{i}%
(\mathbf{q}_{\alpha}\left(  t\right)  ,t)$. Then the momentum conservation law
takes the form
\begin{equation}
D_{t}u_{i}({\mathbf{r}},t)+\partial_{j}\left\langle t_{0ij}({\mathbf{r}%
},t)\right\rangle =-\left\langle n({\mathbf{r}},t)\right\rangle \partial
_{i}v^{{\mathrm{ext}}}({\mathbf{r}},t)\,, \label{3.5}%
\end{equation}
with the material derivative $D_{t}=\partial_{t}+\mathbf{u}({\mathbf{r}%
},t)\cdot\nabla$. In this way the purely convective contributions have been
made explicit.

It remains to calculate the rest frame momentum flux $\left\langle
t_{0ij}({\mathbf{r}},t)\right\rangle $. To do so, the solution to the
Liouville - von Neumann equation is separated into a reference \emph{local
equilibrium} state, $\rho_{N}^{\ell},$ and its remainder $\Delta_{N}$%
\begin{equation}
\rho_{N}\left(  t\right)  =\rho_{N}^{\ell}\left[  y\left(  t\right)  \right]
+\Delta_{N}\left(  t\right)  . \label{3.6}%
\end{equation}
The reference local equilibrium state is chosen to be entirely\ determined by
a set of conjugate fields $\left\{  y(t)\right\}  $ in one-to-one
correspondence with the macroscopic conserved fields $\left\{  \left\langle
n({\mathbf{r}},t)\right\rangle ,\left\langle e({\mathbf{r}},t)\right\rangle
,\left\langle \mathbf{p}({\mathbf{r}},t)\right\rangle \right\}  $. This
correspondence is defined by the requirements that the reference state yield
the exact averages for the local conserved fields
\begin{subequations}
\begin{align}
\overline{n}^{\ell}({\mathbf{r}}|y(t))  &  \equiv\left\langle n({\mathbf{r}%
},t)\right\rangle ,\label{3.7}\\
\overline{e}^{\ell}({\mathbf{r}}|y(t))  &  \equiv\left\langle e({\mathbf{r}%
},t)\right\rangle \,,\label{3.8}\\
\overline{{\mathbf{p}}}^{\ell}({\mathbf{r}}|y(t))  &  \equiv\left\langle
\mathbf{p}({\mathbf{r}},t)\right\rangle , \label{3.9}%
\end{align}
where the superscript $\ell$ denotes a reference ensemble average,
$\overline{A}^{\ell}=\langle A;\rho^{\ell}\rangle$. The left sides of these
equations are functionals of the conjugate fields while the right sides are
the fields of the local conservation laws. In this way the conjugate fields
$\left\{  y(t)\right\}  $ are functionals of the average conserved fields, and
vice versa by inversion. The reference state therefore has the exact average
values for the conserved fields by construction.

A choice for $\rho_{N}^{\ell}$ with these properties is the local equilibrium
ensemble \cite{Zubarev,McLennan}%
\end{subequations}
\begin{equation}
\rho_{N}^{\ell}[y(t)]=e^{-\eta\lbrack y(t)]},\quad\eta\lbrack y(t)]=Q^{\ell
}[y(t)]+\int{\!\!d}\mathbf{r}\psi_{\kappa}({\mathbf{r}})y_{\kappa}%
({\mathbf{r}},t), \label{3.10}%
\end{equation}%
\begin{equation}
Q^{\ell}[y(t)]=\ln\sum_{N}Tr^{(N)}e^{-\int{\!\!d}\mathbf{r}\psi_{\kappa
}({\mathbf{r}})y_{\kappa}({\mathbf{r}},t)} \label{3.10a}%
\end{equation}
where $\psi_{\kappa}({\mathbf{r}})$ are the operators representing the local
conserved\ number density, energy density, and momentum density
\begin{equation}
\left\{  \psi_{\kappa}({\mathbf{r}})\right\}  \equiv\left\{  n({\mathbf{r}%
}),e({\mathbf{r}}),{\mathbf{p}}({\mathbf{r}})\right\}  , \label{3.11}%
\end{equation}
and $y_{\kappa}({\mathbf{r}},t)$ are the conjugate fields,
\begin{equation}
\left\{  y({\mathbf{r}},t)\right\}  \leftrightarrow\left\{  \left[
-\nu({\mathbf{r}},t)+\frac{\beta({\mathbf{r}},t)}{2}mu^{2}({\mathbf{r}%
},t)\right]  ,\beta({\mathbf{r}},t),-\beta({\mathbf{r}},t)\mathbf{u}%
({\mathbf{r}},t)\right\}  \,. \label{3.12}%
\end{equation}
It is interesting to note that this local equilibrium ensemble is also the
\textquotedblleft best choice" in the sense that it maximizes the information
entropy for the given values of the conservative fields
\cite{Jaynes57a,Jaynes57b,Robertson}. In Ref. \citenum{Zubarev} it is also
called the ``relevant" ensemble.

Accordingly, the average rest frame momentum flux $\left\langle t_{0ij}%
({\mathbf{r}},t)\right\rangle $ of (\ref{3.4}) has two contributions. One is
from the reference local equilibrium ensemble $\rho_{N}^{\ell}$ and one from
the remainder in (\ref{3.6})
\begin{align}
\left\langle t_{0ij}({\mathbf{r}},t)\right\rangle  &  =\sum_{N}Tr^{(N)}%
t_{0ij}({\mathbf{r}})\rho_{N}^{\ell}[y(t)]+\sum_{N}Tr^{(N)}t_{0ij}%
({\mathbf{r}})\Delta_{N}\left(  t\right) \nonumber\\
&  \equiv\overline{t_{0ij}}^{\ell}\left(  \mathbf{r}\mid y(t)\right)
+\delta\left\langle t_{0ij}({\mathbf{r}},t)\right\rangle \label{3.13}%
\end{align}
It is shown elsewhere \cite{Dufty20,Zubarev} that the second term of
(\ref{3.13}) describes the dissipative processes of the system while the first
term characterizes the ``perfect fluid" (e.g. Euler) dynamics. The latter is
entirely determined by its functional dependence on the conjugate fields and
reduces to the equilibrium pressure tensor of the last section in the case of
uniform $\beta({\mathbf{r}},t)$. In the following attention will be restricted
to $\overline{t_{0ij}}^{\ell}\left(  \mathbf{r}\mid y(t)\right)  $ and its
possible relationship to an underlying ``local equilibrium thermodynamics."

For the purpose of calculating averages in the rest frame, the macroscopic
velocity dependence can be eliminated so the conjugate fields simplify to
\begin{equation}
\left\{  y({\mathbf{r}})\right\}  \leftrightarrow\left\{  -\nu({\mathbf{r}%
}),\beta({\mathbf{r}}),0\right\}  \,. \label{3.15}%
\end{equation}
Also, the dependence of these fields on time has been suppressed for
simplicity here and below. The hydrodynamic (mechanical) pressure and pressure
tensor are defined in terms of the momentum flux as in the previous section%
\begin{equation}
p_{ij}^{\ell}\left(  \mathbf{r}\mid\beta,\nu\right)  \equiv\overline{t_{0ij}%
}^{\ell}\left(  \mathbf{r}\mid\beta,\nu\right)  , \label{3.16}%
\end{equation}%
\begin{equation}
p_{m}^{\ell}\left(  \mathbf{r}\mid\beta,\nu\right)  \equiv\frac{1}{3}%
\overline{t_{0ii}}^{\ell}\left(  \mathbf{r}\mid\beta,\nu\right)  .
\label{3.17}%
\end{equation}
Since the functional form of the operator $t_{ij}\left(  {\mathbf{r,}%
}t\right)  $ in (\ref{3.14}) is the same for both equilibrium and
non-equilibrium states, the local pressure here is still given by
(\ref{2.20d}) with only the definition of the average changed
\begin{equation}
p_{m}^{\ell}\left(  \mathbf{r}\mid\beta,\nu\right)  =\frac{1}{3}\left(
2\overline{K({\mathbf{r}})}^{\ell}+\overline{\mathcal{V}({\mathbf{r}})}^{\ell
}\right)  . \label{3.18}%
\end{equation}

Next, a thermodynamics associated with the local equilibrium state is defined
in analogy to strict equilibrium via the normalization function $Q^{\ell}[y]$
of (\ref{3.10a}). As noted above the average velocity field can be transformed
to zero so the energy $e({\mathbf{r}})$ is also in the local rest frame. Then
(\ref{3.10a}) becomes
\begin{equation}
Q^{\ell}[\beta,\nu]=\ln\sum_{N}Tr^{(N)}e^{-\int{\!\!d}\mathbf{r}\left(
\beta({\mathbf{r}})e({\mathbf{r}})-\nu({\mathbf{r}})n({\mathbf{r}})\right)  }.
\label{3.14}%
\end{equation}
In analogy to (\ref{2.7}) and (\ref{2.8}), a local equilibrium pressure can be
defined by%
\begin{equation}
\int d\mathbf{r}\beta\left(  \mathbf{r}\right)  p^{\ell}\left(  \mathbf{r}%
\mid\beta,\nu\right)  \equiv Q^{\ell}[\beta,\nu], \label{3.19}%
\end{equation}
or for extensive systems
\begin{equation}
\frac{1}{V}\int d\mathbf{r}\beta\left(  \mathbf{r}\right)  p^{\ell}\left(
\mathbf{r}\mid\beta,\nu\right)  =\left.  \frac{\partial Q^{\ell}[\beta,\nu
]}{\partial V}\right\vert _{\beta,\nu}. \label{3.20}%
\end{equation}
Carrying our the volume derivative leads to
\begin{align}
\left.  \frac{\partial Q^{\ell}[\beta,\nu]}{\partial V}\right\vert _{\beta
,\nu}  &  =\frac{1}{3V}\left.  \overline{\sum_{\alpha=1}^{N}\frac{1}%
{2m}\left[  p_{\alpha j}^{2},\beta({\mathbf{q}}_{\alpha})\right]  _{+}}^{\ell
}\right\vert _{y(t)}\nonumber\\
&  \qquad+\frac{1}{2}\left.  \overline{\sum_{\alpha\neq\gamma=1}^{N}%
\beta({\mathbf{q}}_{\alpha})({\mathbf{q}}_{\gamma}-{\mathbf{q}}_{\alpha}%
)\cdot\mathbf{F}_{\alpha\gamma}(\left\vert {\mathbf{q}}_{\alpha}-{\mathbf{q}%
}_{\gamma}\right\vert )}^{\ell}\right\vert _{y(t)}\label{3.21}\\
&  =\frac{1}{V}\int d\mathbf{r}\beta\left(  \mathbf{r}\right)  \frac{1}%
{3}\left[  2\overline{K_{0}\left(  \mathbf{r}\right)  }^{\ell}+\overline
{\mathcal{V}_{0}\left(  \mathbf{r}\right)  }^{\ell}\right]  \label{3.22}%
\end{align}
where $K_{0}\left(  \mathbf{r}\right)  $ and $\mathcal{V}_{0}\left(
\mathbf{r}\right)  $ are given by (\ref{2.10b}) and (\ref{2.10c}). Therefore,
(\ref{3.20}), or equivalently (\ref{3.19}) gives the identification%
\begin{equation}
Q^{\ell}[\beta,\nu]=\int d\mathbf{r}\beta\left(  \mathbf{r}\right)
p_{0}^{\ell}\left(  \mathbf{r}\mid\beta,\nu\right)  , \label{3.22a}%
\end{equation}
with
\begin{equation}
p_{0}^{\ell}\left(  \mathbf{r}\mid\beta,\nu\right)  =\frac{1}{3}\left[
2\overline{K_{0}\left(  \mathbf{r}\right)  }^{\ell}+\overline{\mathcal{V}%
_{0}\left(  \mathbf{r}\right)  }^{\ell}\right]  . \label{3.23}%
\end{equation}
This is not the same as the hydrodynamic pressure of (\ref{3.18}), $p^{\ell
}\left(  \mathbf{r},\beta\mid\nu\right)  .$ Their volume integrals are the
same%
\begin{equation}
P^{\ell}\left[  \beta,\nu\right]  \equiv\frac{1}{V}\int d\mathbf{r}p_{0}%
^{\ell}\left(  \mathbf{r}\mid\beta,\nu\right)  =\frac{1}{V}\int d\mathbf{r}%
p_{m}^{\ell}\left(  \mathbf{r}\mid\beta,\nu\right)  \label{3.23a}%
\end{equation}
but only $p_{0}^{\ell}\left(  \mathbf{r}\mid\beta,\nu\right)  $ provides the
local density for the grand potential%
\begin{equation}
Q^{\ell}[\beta,\nu]=\int d\mathbf{r}\beta\left(  \mathbf{r}\right)
p_{0}^{\ell}\left(  \mathbf{r}\mid\beta,\nu\right)  \neq\int d\mathbf{r}%
\beta\left(  \mathbf{r}\right)  p_{m}^{\ell}\left(  \mathbf{r}\mid\beta
,\nu\right)  . \label{3.23b}%
\end{equation}

This can be stated in an equivalent way. Using (\ref{2.24}) and (\ref{2.25})
the relationship is%
\begin{equation}
p_{m}^{\ell}\left(  \mathbf{r}\mid\beta,\nu\right)  =p_{0}^{\ell}\left(
\mathbf{r}\mid\beta,\nu\right)  +\Delta p_{0}^{\ell}\left(  \mathbf{r}%
\mid\beta,\nu\right)  \label{3.24}%
\end{equation}

\begin{equation}
\Delta p_{0}^{\ell}\left(  \mathbf{r}\mid\beta,\nu\right)  =\frac{\hbar^{2}%
}{12m}\nabla^{2}\overline{n}^{\ell}\left(  \mathbf{r}\right)  +\frac{1}%
{6}\overline{\sum_{\alpha\neq\sigma=1}^{N}F_{\alpha\sigma i}\left(  \left\vert
\mathbf{q}_{\alpha}-\mathbf{q}_{\sigma}\right\vert \right)  \left[
\mathcal{D}_{i}\left(  \mathbf{r,q}_{\alpha},{\mathbf{q}}_{\sigma}\right)
-({\mathbf{q}}_{\sigma}-{\mathbf{q}}_{\alpha})_{i}\delta\left(  \mathbf{r-q}%
_{\alpha}\right)  \right]  }^{\ell} \,. \label{3.25}%
\end{equation}
Planck's constant has been restored in (\ref{3.25}) to make explicit the fact
that this term has a purely quantum origin. The local pressures differ for
strongly inhomogeneous states, but are the same globally%
\begin{equation}
\int d\mathbf{r}\Delta p_{0}^{\ell}\left(  \mathbf{r}\mid\beta,\nu\right)  =0.
\label{3.26}%
\end{equation}
However, in contrast to the strict equilibrium case of the last section they
do not both give the grand potential due to the appearance of $\beta
({\mathbf{q}}_{\alpha})$ in (\ref{3.21})%
\begin{equation}
\int d\mathbf{r}\beta\left(  \mathbf{r}\right)  \Delta p_{0}^{\ell}\left(
\mathbf{r}\mid\beta,\nu\right)  \neq0. \label{3.27}%
\end{equation}
Thus the hydrodynamic local pressure, $p_{m}^{\ell}\left(  \mathbf{r}\mid
\beta,\nu\right)  $, does not have the expected relationship to thermodynamics.

The seeming paradox now is that the stationary solution to the hydrodynamic
equations is (\ref{2.29})%
\begin{equation}
\partial_{j}p_{ij}^{\ell}\left(  \mathbf{r},\beta\mid\nu\right)
=-\left\langle n({\mathbf{r}})\right\rangle \partial_{i}v^{{\mathrm{ext}}%
}({\mathbf{r}}), \label{3.28}%
\end{equation}%
\begin{equation}
p_{ij}^{\ell}\left(  \mathbf{r},\beta\mid\nu\right)  =\frac{1}{3}p_{m}^{\ell
}\left(  \mathbf{r},\beta\mid\nu\right)  \delta_{ij}+\widetilde{p}_{ij}^{\ell
}\left(  \mathbf{r},\beta\mid\nu\right)  , \label{3.29}%
\end{equation}
but the corresponding pressure $p_{m}^{\ell}\left(  \mathbf{r},\beta\mid
\nu\right)  $ is not the thermodynamic local pressure\ $p_{0}^{\ell}\left(
\mathbf{r}\mid\beta,\nu\right)  $ even in the limit of uniform $\beta$. The
hydrodynamic stationary state is not the thermodynamic state.

\section{Revised momentum flux and hydrodynamic pressure}

While there is no flexibility in the choice $p_{0}^{\ell}\left(
\mathbf{r}\mid\beta,\nu\right)  $ in the first equality of (\ref{3.23b}) due
to the space dependence of $\beta\left(  \mathbf{r}\right)  $ there is some
ambiguity in $p_{m}^{\ell}\left(  \mathbf{r}\mid\beta,\nu\right)  $ available
to remove the inequality. This is due to the fact that only the space
derivative of $t_{0ij}({\mathbf{r}})$ occurs in (\ref{3.5}). Consequently, any
tensor of the form%
\begin{equation}
\overline{t_{0ij}^{\prime}}^{\ell}({\mathbf{r}})=\overline{t_{0ij}}^{\ell
}({\mathbf{r}})+\epsilon_{jkn}\partial_{k}A_{ni}({\mathbf{r}}) \label{4.1}%
\end{equation}
will give an equivalent derivative.\ \ More generally, it is shown that the
contour in the definition of $\mathcal{D}_{j}\left(  \mathbf{r,q}%
_{1},\mathbf{q}_{2}\right)  $, (\ref{2.15}) can be chosen such that
$t_{0ij}({\mathbf{r}})=t_{0ji}({\mathbf{r}})$ (see Supporting Information S.2), and
hence%
\begin{equation}
\partial_{j}\overline{t_{0ij}}^{\ell}({\mathbf{r}})=\partial_{j}%
\overline{t_{0ji}}^{\ell}({\mathbf{r}}). \label{4.2}%
\end{equation}
Then the generalization of (\ref{4.1}) to its symmetric form is%
\begin{equation}
\overline{t_{0ij}^{\prime}}^{\ell}=\overline{t_{0ij}}^{\ell}+\epsilon_{ik\ell
}\epsilon_{jmn}\partial_{k}\partial_{m}A_{\ell n},\hspace{0.2in}A_{\ell
n}=A_{n\ell} \,, \label{4.3}%
\end{equation}
where here $\overline{t_{0ij}}^{\ell}$ is the symmetric form of Supporting
Information S.2. It is readily verified that
\begin{equation}
\partial_{j}\overline{t_{0ij}^{\prime}}^{\ell}({\mathbf{r}})=\partial
_{j}\overline{t_{0ij}}^{\ell}({\mathbf{r}})\hspace{0.2in}\partial_{i}%
\overline{t_{0ij}^{\prime}}^{\ell}({\mathbf{r}})=\partial_{i}\overline
{t_{0ij}}^{\ell}({\mathbf{r}}) \,. \label{4.4}%
\end{equation}

The hydrodynamic pressure tensor associated with $\overline{t_{0ij}^{\prime}}^\ell%
({\mathbf{r}},t)$ is%

\begin{align}
p_{ij}^{\prime\ell}\left(  \mathbf{r}\mid\beta,\nu\right)   &  \equiv
\overline{t_{0ij}^{\prime}}^{\ell}\left(  \mathbf{r}\mid\beta,\nu\right)
=\overline{t_{0ij}}^{\ell}\left(  \mathbf{r}\mid\beta,\nu\right)
+\epsilon_{ik\ell}\epsilon_{jmn}\partial_{k}\partial_{m}A_{\ell n}%
({\mathbf{r}}\mid\beta,\nu)\nonumber\\
&  =p_{ij}^{\ell}\left(  \mathbf{r}\mid\beta,\nu\right)  +\epsilon_{ik\ell
}\epsilon_{jmn}\partial_{k}\partial_{m}A_{\ell n}({\mathbf{r}}\mid\beta,\nu)
\label{4.5}%
\end{align}
and the corresponding pressure is
\begin{align}
p^{\prime\ell}\left(  \mathbf{r}\mid\beta,\nu\right)   &  =p_{m}^{\ell}\left(
\mathbf{r}\mid\beta,\nu\right)  +\frac{1}{3}\epsilon_{ik\ell}\epsilon
_{imn}\partial_{k}\partial_{m}A_{\ell n}({\mathbf{r}}\mid\beta,\nu)\nonumber\\
&  =p_{m}^{\ell}\left(  \mathbf{r}\mid\beta,\nu\right)  +\frac{1}{3}\left(
\partial_{k}^{2}A_{\ell\ell}({\mathbf{r}}\mid\beta,\nu)-\partial_{k}%
\partial_{\ell}A_{\ell k}({\mathbf{r}}\mid\beta,\nu)\right)  . \label{4.6}%
\end{align}
Use has been made of the identity%
\begin{equation}
\epsilon_{ik\ell}\epsilon_{imn}=\delta_{km}\delta_{\ell n}-\delta_{kn}%
\delta_{\ell m}. \label{4.7}%
\end{equation}
To further simplify (\ref{4.6}) choose the arbitrary tensor $A_{\ell
k}({\mathbf{r}}\mid\beta,\nu)$ to be diagonal%
\begin{equation}
A_{\ell k}({\mathbf{r}}\mid\beta,\nu)\rightarrow\delta_{k\ell}A({\mathbf{r}%
}\mid\beta,\nu), \label{4.8}%
\end{equation}
giving%
\begin{equation}
p^{\prime\ell}\left(  \mathbf{r}\mid\beta,\nu\right)  =p^{\ell}\left(
\mathbf{r}\mid\beta,\nu\right)  +\frac{2}{3}\nabla^{2}A({\mathbf{r}}\mid
\beta,\nu)\,, \label{4.9}%
\end{equation}%
\begin{equation}
p_{ij}^{\prime\ell}\left(  \mathbf{r}\mid\beta,\nu\right)  =p_{ij}^{\ell
}\left(  \mathbf{r}\mid\beta,\nu\right)  +\left(  \delta_{ij}\nabla
^{2}-\partial_{j}\partial_{i}\right)  A({\mathbf{r}}\mid\beta,\nu) \,.
\label{4.9a}%
\end{equation}

It is now seen that the hydrodynamic pressure $p^{\prime\ell}\left(
\mathbf{r}\mid\beta,\nu\right)  $ can be equated to the thermodynamic pressure
$p_{0}^{\ell}\left(  \mathbf{r}\mid\beta,\nu\right)  $ by the \ choice%

\begin{equation}
\frac{2}{3}\nabla^{2}A({\mathbf{r}}\mid\beta,\nu)=-\Delta p_{0}^{\ell}\left(
\mathbf{r}\mid\beta,\nu\right) \,.  \label{4.10}%
\end{equation}
Since $\Delta p_{0}^{\ell}\left(  \mathbf{r}\mid\beta,\nu\right)  $ is given
explicitly by (\ref{3.25}), this Poisson's equation for $A({\mathbf{r}}%
\mid\beta,\nu)$ is well-defined. In summary, by modifying the form for the
average stress tensor the momentum balance equation is unchanged but the
mechanical pressure can be chosen equal to the equilibrium pressure for both
the equilibrium and local equilibrium states, and that pressure is
$p_{0}^{\ell}\left(  \mathbf{r}\mid\beta,\nu\right)  $ or $p_{0}^{e}\left(
\mathbf{r,}\beta\mid\nu\right)  $ respectively.

As a special simple case the explicit results for a non-interacting
inhomogeneous gas are
\begin{equation}
\frac{2}{3}\nabla^{2}A({\mathbf{r}}\mid\beta,\nu)=-\frac{\hbar^{2}}{12m}%
\nabla^{2}\overline{n}^{\ell}\left(  \mathbf{r}\right)  , \label{4.11}%
\end{equation}
with the solution%
\begin{equation}
A({\mathbf{r}}\mid\beta,\nu)=-\frac{\hbar^{2}}{8m}\overline{n}^{\ell}\left(
\mathbf{r}\right)  +a({\mathbf{r}}\mid\beta,\nu),\hspace{0.2in}\nabla
^{2}a({\mathbf{r}}\mid\beta,\nu)=0. \label{4.12}%
\end{equation}

In summary, by changing the form of the pressure tensor to (\ref{4.5}) there
is the freedom to choose the new mechanical local pressure $p^{\prime\ell
}\left(  \mathbf{r}\mid\beta,\nu\right)  $ to be equal to the thermodynamic
local pressure $p_{0}^{\ell}\left(  \mathbf{r}\mid\beta,\nu\right)  $. In this
way the desired equality%
\begin{equation}
Q^{\ell}[\beta,\nu]=\int d\mathbf{r}\beta\left(  \mathbf{r}\right)
p_{0}^{\ell}\left(  \mathbf{r}\mid\beta,\nu\right)  =\int d\mathbf{r}%
\beta\left(  \mathbf{r}\right)  p_{m}^{\prime\ell}\left(  \mathbf{r}\mid
\beta,\nu\right)  , \label{4.13}%
\end{equation}
is recovered. This result holds for both equilibrium and local equilibrium states.

\section{Discussion}

The definition of a local pressure from two conceptually different origins has
been considered. The first is thermodynamic in nature, associated with the
local density for the grand potential. The other is mechanical in nature,
associated with the average of the local momentum flux (referred to as the
mechanical or hydrodynamic pressure). First, a strictly equilibrium state for
an inhomogeneous system was described. This is the case of interest for
density functional theory where the task is to calculate the global free
energy. Although local free energy densities are introduced in that context
they are mainly for computational convenience. However, they also provide the
basis for a local thermodynamics as well. For example, a local pressure
follows from a local ``Legendre transformation" of the free energy density
\cite{Gubbins93}. In the present case, the thermodynamic local pressure is
introduced directly as the density for the grand potential. It is then
compared to the pressure defined from the equilibrium average of the momentum
flux (equilibrium force balance equation). The two are different for strongly
inhomogeneous states. However, the possibility of adding a contribution
$\Delta p_{0}^{\ell}\left(  \mathbf{r}\mid\beta,\nu\right)  $ whose volume
integral vanishes to the grand potential density can be exploited to assure
the thermodynamic and hydrodynamic local pressures are the same.

The same analysis for \textit{local equilibrium} states, where the temperature
is also non-uniform, again leads to different forms for the thermodynamic and
hydrodynamic local pressures. But in this case there is no longer the
flexibility to add a contribution $\Delta p_{0}^{\ell}\left(  \mathbf{r}%
\mid\beta,\nu\right)  $ to the thermodynamic local pressure for resolution.
Instead, the hydrodynamic local pressure can be changed by exploiting the fact
that the momentum flux occurs in the momentum conservation law only as a
divergence of that flux. In this way agreement of the thermodynamic and
hydrodynamic local pressures is restored.

For consistency between the local and strict equilibrium states in the limit
of uniform temperature, the local equilibrium form must be used, $p_{0}^{\ell
}\left(  \mathbf{r}\mid\beta,\nu\right)  =p_{m}^{\prime\ell}\left(
\mathbf{r}\mid\beta,\nu\right)  $, i.e. the same average momentum flux should
be adopted in each case. This is a somewhat simpler form than the pressure
tensor and local pressure than $p_{m}^{\ell}\left(  \mathbf{r}\mid\beta
,\nu\right)  $ for the strict equilibrium case. The requirement that the force
balance equation at local equilibrium gives the correct local density for the
local equilibrium grand potential gives a strong constraint on the equilibrium
form as well. 

As noted in (\ref{3.23a}) both $p_{0}^{\ell}\left(  \mathbf{r}\mid\beta
,\nu\right)  $ and $p_{m}^{\ell}\left(  \mathbf{r}\mid\beta,\nu\right)  $ are
the local densities for the global pressure, without any need for modification
of the pressure tensor. If that were chosen to be the constraint of
thermodynamic consistency, then no modification of the choice $p_{m}^{\ell
}\left(  \mathbf{r}\mid\beta,\nu\right)  =p^{\ell}\left(  \mathbf{r}\mid
\beta,\nu\right)  $ as in the equilibrium case is required - no change in the
pressure tensor. Agreement of the equilibrium and local equilibrium cases
would be direct, but the inequality (\ref{3.23b}) would remain. The
consequences of this for local equilibrium thermodynamics is not clear and
needs to be explored further. The grand potential is a Massieu-Planck
functional in the foundations of local equilibrium thermodynamics
\cite{Zubarev}. The pressure functional alone has no corresponding role.

Reference has been made above to density functional theory where the pressure
is expressed as a functional of the density $n$ rather than the activity $\nu
$. This change of variables is obtained by inverting
\begin{equation}
\overline{n({\mathbf{r}}\mid\beta,\nu)}^{\ell}=\overline{n}^{\ell}%
\hspace{0.2in}\rightarrow\hspace{0.2in}\nu=\nu({\mathbf{r}}\mid\beta,n).
\label{5.1}%
\end{equation}
This is a difficult problem, separate from the discussion referring to the
definition of $p_{0}^{\ell}\left(  \mathbf{r}\mid\beta,\nu\right)  $.
\ However, the proof that the hydrodynamic pressure can be chosen to be the
same as the thermodynamic pressure assures that the tools of density
functional theory can be used within the hydrodynamic context as well.

In closing it is useful to return to the extensive literature mentioned above regarding the
definition of the pressure tensor and its measurement by molecular dynamics simulation\cite{Hardy82,Walton83,Todd95,Heyes11}, and
its common implementation in the Sandia National Laboratories code LAMMPS (Large-scale
Atomic/Molecular Massively Parallel Simulator). These simulations refer to methods for direct
evaluation of the microscopic definitions of the various components for the pressure tensor. Here,
no consideration is given for the pressure tensor components beyond its scalar trace. In that respect
the equivalence constraint does not determine the full pressure tensor. Also, only its local
equilibrium average is involved --- the residual irreversible component of the momentum flux is not
affected. A direct measurement of the momentum flux by simulation would give the total of both
components (see eq.(\ref{3.13})), which would give the part studied here only for non-dissipative flows, e.g.,
the equilibrium state. Finally, the equivalence condition is the equality of two functionals of the
density and temperature. The numerical confirmation of functional equivalence is indeed a
formidable task. Thus the important simulation studies are only of indirect bearing on the limited
scope considered here.

\section{Acknowledgments}

The authors are indebted to S.B. Trickey for comments and criticism of
early drafts. This research was supported by US DOE Grant DE-SC0002139.  K.L. was
also supported by U.S. NSF CSEDI grant EAR-1901813 and the Carnegie Institution
for Science.

\end{document}


\title{Supporting Information for\\
 \bigskip
Local pressure for inhomogeneous fluids}

\author{James Dufty}
\email{dufty@ufl.edu}
\affiliation{Department of Physics, University of Florida, Gainesville, FL
32611, USA}
\author{Jeffrey Wrighton}
\email{wrighton@ufl.edu}
\affiliation{Department of Physics, University of Florida, Gainesville, FL
32611, USA}
\author{Kai Luo}
\email{kluo@carnegiescience.edu}
\affiliation{Earth and Planets Laboratory, Carnegie Institution for Science,
Washington, DC 20015-1305, USA}

\renewcommand{\theequation}{S.\arabic{equation}}
\renewcommand{\thesection}{S.\arabic{section}}

\setcitestyle{super}

\renewcommand{\bibnumfmt}[1]{[S#1]}
\renewcommand{\citenumfont}[1]{S#1}

\date{\today}
\maketitle

\section{Conservation of number and momentum densities}

\label{ap:A}The dynamics of operators is defined by%
\begin{equation}
\partial_{t}X\left(  t\right)  =i\left[  \mathcal{H}_{N}\left(  t\right)
,X\left(  t\right)  \right]  , \label{a.1a}%
\end{equation}
with the Hamiltonian given by 

\begin{equation}
\mathcal{H}_{N}\left(  t\right)  =H_{N}\left(  t\right)  +\sum_{\alpha=1}%
^{N}v^{{\mathrm{ext}}}({\mathbf{q}}_{\alpha}\left(  t\right)  ,t)),
\label{a.2}%
\end{equation}%
\begin{equation}
H_{N}\left(  t\right)  =\sum_{\alpha=1}^{N}\frac{p_{\alpha}^{2}\left(
t\right)  }{2m}+\frac{1}{2}\sum_{\alpha\neq\sigma=1}^{N}U_{N}(\left\vert
{\mathbf{q}}_{\alpha}\left(  t\right)  -{\mathbf{q}}_{\sigma}\left(  t\right)
\right\vert ). \label{a.3}%
\end{equation}
The operators corresponding to the number and momentum densities are%
\begin{equation}
n\left(  \mathbf{r},t\right)  =\sum_{\alpha=1}^{N}\delta\left(  \mathbf{r-q}%
_{\alpha}\left(  t\right)  \right)  ,\hspace{0.24in}\mathbf{p}\left(
\mathbf{r},t\right)  =\sum_{\alpha=1}^{N}\frac{1}{2}\left[  \mathbf{p}%
_{\alpha}\left(  t\right)  ,\delta\left(  \mathbf{r-q}_{\alpha}\left(
t\right)  \right)  \right]  _{+}. \label{a.3a}%
\end{equation}
The brackets $\left[  A,B\right]  _{+}$with a subscript $+$ denote an anti-commutator.

Local conservation laws follow exactly from this Hamiltonian dynamics
\cite{Dufty20}. The simplest is the conservation of number density%
\begin{align}
\partial_{t}n\left(  \mathbf{r,}t\right)   &  =i\left[  \mathcal{H}_{N}\left(
t\right)  ,n\left(  \mathbf{r,}t\right)  \right]  =\sum_{\alpha=1}^{N}i\left[
\frac{p_{\alpha}^{2}\left(  t\right)  }{2m},\delta\left(  \mathbf{r-q}%
_{\alpha}\left(  t\right)  \right)  \right] \nonumber\\
&  =\sum_{\alpha=1}^{N}i\left\{  \frac{p_{\alpha j}\left(  t\right)  }%
{2m}\left[  p_{\alpha j}\left(  t\right)  ,\delta\left(  \mathbf{r-q}_{\alpha
}\left(  t\right)  \right)  \right]  +\left[  p_{\alpha j}\left(  t\right)
,\delta\left(  \mathbf{r-q}_{\alpha}\left(  t\right)  \right)  \right]
\frac{p_{\alpha j}\left(  t\right)  }{2m}\right\} \nonumber\\
&  =\frac{1}{2m}\sum_{\alpha=1}^{N}\left\{  p_{\alpha j}\left(  t\right)
\partial_{q_{\alpha j}\left(  t\right)  }\delta\left(  \mathbf{r-q}_{\alpha
}\left(  t\right)  \right)  +\partial_{q_{\alpha j}\left(  t\right)  }%
\delta\left(  \mathbf{r-q}_{\alpha}\left(  t\right)  \right)  p_{\alpha
j}\left(  t\right)  \right\}  \label{a.4}%
\end{align}
or, with the definition of $\mathbf{p}\left(  \mathbf{r}\right)  $ in
(\ref{a.3a}), the microscopic continuity equation is obtained
\begin{equation}
\partial_{t}n\left(  \mathbf{r,}t\right)  +\frac{1}{m}\nabla\cdot
\mathbf{p}\left(  \mathbf{r},t\right)  =0. \label{a.5}%
\end{equation}
Use has been made of the property%
\begin{equation}
i\left[  p_{\alpha j}\left(  t\right)  ,A\left(  \mathbf{q}_{\alpha}\left(
t\right)  \right)  \right]  =\partial_{q_{\alpha j}\left(  t\right)  }A\left(
\mathbf{q}_{\alpha}\left(  t\right)  \right)  . \label{a.5a}%
\end{equation}

The time derivative of $\mathbf{p}\left(  \mathbf{r},t\right)  $ follows in a
similar way%
\begin{align*}
\partial_{t}p_{j}\left(  \mathbf{r},t\right)     = &\frac{1}{2}\sum_{\alpha
=1}^{N}\left\{  \left[  p_{\alpha j}\left(  t\right)  ,i\left[  \mathcal{H}%
_{N}\left(  t\right)  ,\delta\left(  \mathbf{r-q}_{\alpha}\left(  t\right)
\right)  \right]  \right]  _{+}+\left[  i\left[  \mathcal{H}_{N}\left(
t\right)  ,p_{\alpha j}\left(  t\right)  \right]  ,\delta\left(
\mathbf{r-q}_{\alpha}\left(  t\right)  \right)  \right]  _{+}\right\} \\
  = &\frac{1}{2}\sum_{\alpha=1}^{N}\left\{  \left[  p_{\alpha j}\left(
t\right)  ,i\left[  \frac{p_{\alpha}^{2}\left(  t\right)  }{2m},\delta\left(
\mathbf{r-q}_{\alpha}\left(  t\right)  \right)  \right]  \right]  _{+}+\left[
i\left[  \mathcal{H}_{N}\left(  t\right)  ,p_{\alpha j}\left(  t\right)
\right]  ,\delta\left(  \mathbf{r-q}_{\alpha}\left(  t\right)  \right)
\right]  _{+}\right\} \\
  = &\frac{1}{4m}\sum_{\alpha=1}^{N}\left[  p_{\alpha j}\left(  t\right)
,\left[  p_{\alpha k}\left(  t\right)  ,\partial_{q_{\alpha k}\left(
t\right)  }\Delta\left(  \mathbf{r-q}_{\alpha}\left(  t\right)  \right)
\right]  _{+}\right]  _{+}-\frac{1}{2}\sum_{\alpha=1}^{N}\delta\left(
\mathbf{r-q}_{\alpha}\left(  t\right)  \right)  \partial_{q_{\alpha j}\left(
t\right)  }v^{{\mathrm{ext}}}({\mathbf{q}}_{\alpha}\left(  t\right)  ,t))\\
&  -\frac{1}{4}\sum_{\alpha=1}^{N}\sum_{\beta,\sigma=1}^{N}\left[
\delta\left(  \mathbf{r-q}_{\alpha}\left(  t\right)  \right)  ,\partial
_{q_{\alpha j}\left(  t\right)  }V\left(  \left\vert \mathbf{q}_{\beta}\left(
t\right)  -\mathbf{q}_{\sigma}\left(  t\right)  \right\vert \right)  \right]
_{+}%
\end{align*}%
\begin{align}
  =&-\partial_{r_{k}}\frac{1}{4m}\sum_{\alpha=1}^{N}\left[  p_{\alpha
j}\left(  t\right)  ,\left[  p_{\alpha k}\left(  t\right)  ,\delta\left(
\mathbf{r-q}_{\alpha}\left(  t\right)  \right)  \right]  _{+}\right]
_{+}+\frac{1}{2}\sum_{\alpha=1}^{N}\delta\left(  \mathbf{r-q}_{\alpha}\left(
t\right)  \right)  F_{\alpha j}^{{\mathrm{ext}}}\left(  \mathbf{q}_{\alpha
}\left(  t\right)  \right) \nonumber\\
&\  +\frac{1}{2}\sum_{\alpha=1}^{N}\sum_{\beta=1}^{N}F_{\alpha\beta j}\left(
\left\vert \mathbf{q}_{\alpha}\left(  t\right)  -\mathbf{q}_{\beta}\left(
t\right)  \right\vert \right)  \delta\left(  \mathbf{r-q}_{\alpha}\left(
t\right)  \right)  \label{a.6}%
\end{align}
where $F_{\alpha\beta j}\left(  \left\vert \mathbf{q}_{\alpha}\left(
t\right)  -\mathbf{q}_{\beta}\left(  t\right)  \right\vert \right)  $ is the
$j^{th\text{ }}$ component of the force on particle $\alpha$ due to particle
$\beta$%
\begin{equation}
F_{\alpha\beta j}\left(  \left\vert \mathbf{q}_{\alpha}-\mathbf{q}_{\beta
}\right\vert \right)  =-\partial_{q_{\alpha j}\left(  t\right)  }V\left(
\left\vert \mathbf{q}_{\alpha}\left(  t\right)  -\mathbf{q}_{\beta}\left(
t\right)  \right\vert \right)  , \label{a.8}%
\end{equation}
and $F_{\alpha j}^{{\mathrm{ext}}}\left(  \mathbf{q}_{\alpha}\left(  t\right)
\right)  $ is the $j^{th\text{ }}$ component of the force on particle $\alpha$
due to the external potential
\begin{equation}
F_{\alpha j}^{\rm ext}\left(  \mathbf{q}_{j}\right)  =-\partial_{q_{\alpha
j}\left(  t\right)  }u\left(  \mathbf{q}_{\alpha}\left(  t\right)  ,t\right)
. \label{a.10}%
\end{equation}

The last term on the right side of (\ref{a.6}) can be rewritten as
\begin{align}
\sum_{\alpha\neq\beta=1}^{N}F_{\alpha\beta j}\left(  \left\vert \mathbf{q}%
_{\alpha}\left(  t\right)  -\mathbf{q}_{\beta}\left(  t\right)  \right\vert
\right)  \delta\left(  \mathbf{r-q}_{\alpha}\left(  t\right)  \right)
=&\sum_{\alpha\neq\beta=1}^{N}F_{\beta\alpha j}\left(  \left\vert
\mathbf{q}_{\beta}\left(  t\right)  -\mathbf{q}_{\alpha}\left(  t\right)
\right\vert \right)  \delta\left(  \mathbf{r-q}_{\beta}\left(  t\right)
\right)\nonumber\\
=&-\sum_{\alpha\neq\beta=1}^{N}F_{\alpha\beta j}\left(  \left\vert
\mathbf{q}_{\alpha}\left(  t\right)  -\mathbf{q}_{\beta}\left(  t\right)
\right\vert \right)  \delta\left(  \mathbf{r-q}_{\beta}\left(  t\right)
\right)  \label{a.11}%
\end{align}
as follows from Newton's third law. Therefore%
\begin{align}
\sum_{\alpha\neq\beta=1}^{N}F_{\alpha\beta j}\left(  \left\vert \mathbf{q}%
_{\alpha}\left(  t\right)  -\mathbf{q}_{\beta}\left(  t\right)  \right\vert
\right)  \delta\left(  \mathbf{r-q}_{\alpha}\left(  t\right)  \right)   
= &\frac{1}{2}\sum_{\alpha\neq\beta=1}^{N}F_{\alpha\beta j}\left(  \left\vert
\mathbf{q}_{\alpha}\left(  t\right)  -\mathbf{q}_{\beta}\left(  t\right)
\right\vert \right) \nonumber\\
&  \times\left(  \delta\left(  \mathbf{r-q}_{\alpha}\left(  t\right)  \right)
-\delta\left(  \mathbf{r-q}_{\beta}\left(  t\right)  \right)  \right).
\label{a.11a}%
\end{align}
Next note the identity%
\begin{equation}
\delta\left(  \mathbf{r-q}_{1}\right)  -\delta\left(  \mathbf{r-q}_{2}\right)
=\int_{\lambda_{2}}^{\lambda_{1}}d\lambda\frac{d}{d\lambda}\delta\left(
\mathbf{r-x}\left(  \lambda\right)  \right)  ,\hspace{0.26in}\mathbf{x}\left(
\lambda_{1}\right)  =\mathbf{q}_{1},\hspace{0.26in}\mathbf{x}\left(
\lambda_{2}\right)  =\mathbf{q}_{2}. \label{a.13}%
\end{equation}
Here $\mathbf{x}\left(  \lambda\right)  $ is an arbitrary path for the vector
$\mathbf{x}$ moving from $\mathbf{x}\left(  \lambda_{2}\right)  =\mathbf{q}%
_{2}$ to $\mathbf{x}\left(  \lambda_{1}\right)  =\mathbf{q}_{1}$. Carrying out
the derivative gives%
\begin{equation}
\delta\left(  \mathbf{r-q}_{1}\right)  -\delta\left(  \mathbf{r-q}_{2}\right)
=-\partial_{r_{k}}\mathcal{D}_{k}\left(  \mathbf{r,q}_{1},\mathbf{q}%
_{2}\right)  , \label{a.14}%
\end{equation}
where%
\begin{equation}
\mathcal{D}_{k}\left(  \mathbf{r,q}_{1},\mathbf{q}_{2}\right)  =\int
_{\lambda_{2}}^{\lambda_{1}}d\lambda\frac{dx_{k}\left(  \lambda\right)
}{d\lambda}\delta\left(  \mathbf{r-x}\left(  \lambda\right)  \right)  .
\label{a.15}%
\end{equation}
Use of (\ref{a.14}) in (\ref{a.11a}) and (\ref{a.6}) gives the momentum
conservation law%

\begin{equation}
\partial_{t}p_{j}\left(  \mathbf{r},t\right)  +\partial_{r_{k}}t_{jk}\left(
\mathbf{r},t\right)  =f_{j}\left(  \mathbf{r,}t\right)  \label{a.16}%
\end{equation}
where the total momentum flux $t_{\alpha\beta}\left(  \mathbf{r},t\right)  $
is the sum of a kinetic and potential part%
\begin{equation}
t_{\alpha\beta}\left(  \mathbf{r},t\right)  =t_{\alpha\beta}^{K}\left(
\mathbf{r},t\right)  +t_{\alpha\beta}^{P}\left(  \mathbf{r},t\right)  ,
\label{a.17}%
\end{equation}%
\begin{equation}
t_{jk}^{K}\left(  \mathbf{r},t\right)  =\frac{1}{4m}\sum_{\alpha=1}^{N}\left[
p_{\alpha j}\left(  t\right)  ,\left[  p_{\alpha k}\left(  t\right)
,\delta\left(  \mathbf{r-q}_{\alpha}\left(  t\right)  \right)  \right]
_{+}\right]  _{+} ,\label{a.17a}%
\end{equation}%
\begin{equation}
t_{jk}^{P}\left(  \mathbf{r},t\right)  =\frac{1}{4}\sum_{\alpha\neq\beta
=1}^{N}F_{\alpha\beta j}\left(  \left\vert \mathbf{q}_{\alpha}\left(
t\right)  -\mathbf{q}_{\beta}\left(  t\right)  \right\vert \right)
\mathcal{D}_{k}\left(  \mathbf{r,q}_{\alpha}\left(  t\right)  ,\mathbf{q}%
_{\beta}\left(  t\right)  \right).  \label{a.18}%
\end{equation}
The right side of (\ref{a.16}) is the force density of the external potential%
\begin{equation}
f_{j}\left(  \mathbf{r,}t\right)  =\sum_{\alpha=1}^{N}\delta\left(
\mathbf{r-q}_{j}\left(  t\right)  \right)  F_{\alpha j}^{\rm ext}\left(
\mathbf{q}_{\alpha}\left(  t\right)  \right).  \label{a.19}%
\end{equation}

\section{Determination of momentum flux}

\label{ap:B}The straightforward derivation of the momentum conservation law in
Appendix \ref{ap:A} leads to a momentum flux (as in the text the dependence on
$t$ will be left implicit as it plays no role in the following)%
\begin{align}
t_{ij}\left(  {\mathbf{r}}\right)     = &\frac{1}{4m}\sum_{\alpha=1}^{N}\left[
p_{i\alpha},\left[  p_{j\alpha},\delta\left(  \mathbf{r-q}_{\alpha}\right)
\right]  _{+}\right]  _{+}\nonumber\\
&  +\frac{1}{2}\sum_{\alpha\neq\sigma=1}^{N}F_{\alpha\sigma i}\left(
\left\vert \mathbf{q}_{\alpha}-\mathbf{q}_{\sigma}\right\vert \right)
\mathcal{D}_{j}\left(  \mathbf{r,q}_{\alpha},{\mathbf{q}}_{\sigma}\right)
,\label{b.1}%
\end{align}
where the operator $\mathcal{D}_{j}\left(  \mathbf{r,q}_{\alpha},{\mathbf{q}%
}_{\sigma}\right)  $ is given by%
\begin{equation}
\mathcal{D}_{j}\left(  \mathbf{r,q}_{1},\mathbf{q}_{2}\right)  \equiv
\int_{\mathcal{C}}d\lambda\frac{dx_{j}\left(  \lambda\right)  }{d\lambda
}\delta\left(  \mathbf{r-x}\left(  \lambda\right)  \right)  ,\hspace
{0.26in}\mathbf{x}\left(  \lambda_{1}\right)  =\mathbf{q}_{1},\hspace
{0.26in}\mathbf{x}\left(  \lambda_{2}\right)  =\mathbf{q}_{2}.\label{b.2}%
\end{equation}
Here $\mathcal{C}$ is an arbitrary continuous path connecting $\mathbf{x}%
\left(  \lambda\right)  $ between $\lambda_{1}$ and $\lambda_{2}$.
Consider the simplest choice a linear path%
\begin{equation}
\mathbf{x}\left(  \lambda\right)  =\mathbf{q}_{1}+\left(  \mathbf{q}%
_{2}-\mathbf{q}_{1}\right)  \lambda.\label{b.3}%
\end{equation}
Then%
\begin{equation}
\mathcal{D}_{j}\left(  \mathbf{r,q}_{1},\mathbf{q}_{2}\right)  \rightarrow
\left(  q_{2j}-q_{1j}\right)  \int_{0}^{1}d\lambda\delta\left(
\mathbf{r-\mathbf{q}_{1}-\left(  \mathbf{q}_{2}-\mathbf{q}_{1}\right)
}\lambda\right)  ,\hspace{0.26in}\mathbf{x}\left(  \lambda_{1}\right)
=\mathbf{q}_{1},\hspace{0.26in}\mathbf{x}\left(  \lambda_{2}\right)
=\mathbf{q}_{2}.\label{b.4}%
\end{equation}
Since $F_{\alpha\sigma i}\left(  \left\vert \mathbf{q}_{\alpha}-\mathbf{q}%
_{\sigma}\right\vert \right)  \propto\left(  q_{2i}-q_{1i}\right)  $ use of
(\ref{b.4}) in (\ref{b.1}) leads to a form for the momentum flux that is
symmetric, $t_{ij}\left(  {\mathbf{r}}\right)  =t_{ji}\left(  {\mathbf{r}%
}\right)  $, and consequently its divergence with respect to first and second
indices are the same%
\begin{equation}
\partial_{j}t_{ij}\left(  {\mathbf{r}}\right)  =\partial_{j}t_{ji}\left(
{\mathbf{r}}\right)  .\label{b.5}%
\end{equation}

Next define another symmetric momentum flux whose divergences are the same as
those of (\ref{b.5})%

\begin{equation}
t_{ij}^{\prime}\left(  {\mathbf{r}}\right)  =t_{ij}\left(  {\mathbf{r}%
}\right)  +\epsilon_{ik\ell}\epsilon_{jmn}\partial_{k}\partial_{m}A_{\ell
n}\left(  {\mathbf{r}}\right)  , \label{b.6}%
\end{equation}
for some unspecified tensor field $A_{\ell n}\left(  {\mathbf{r}}\right)  $
and $\epsilon_{ik\ell}$ is the Levi-Cevita tensor. This added term is seen to
be the curl of a vector associated with each component of $t_{ij}\left(
{\mathbf{r}}\right)  $ (i.e., noting that $t_{ij}\left(  {\mathbf{r}}\right)
$ transforms as vector for components $i$ at fixed $j$, and vice versa). With
the additional condition $A_{\ell n}\left(  {\mathbf{r}}\right)  =A_{n\ell
}\left(  {\mathbf{r}}\right)  $, it is verified that%
\begin{equation}
t_{ij}^{\prime}\left(  {\mathbf{r}}\right)  =t_{ji}^{\prime}\left(
{\mathbf{r}}\right)\,,  \label{b.7}%
\end{equation}%
\begin{equation}
\partial_{j}t_{ij}^{\prime}\left(  {\mathbf{r}}\right)  =\partial_{j}%
t_{ji}^{\prime}\left(  {\mathbf{r}}\right)  . \label{b.8}%
\end{equation}

The two indefinite features of the momentum flux have now been made explicit
with the contour fixed by symmetry and the choice of a term with vanishing
divergence. It is interesting to note that the contribution from different
contours, denoted by $\Delta\mathcal{D}_{j}\left(  \mathbf{r,q}_{1}%
,\mathbf{q}_{2}\right)  $ has a vanishing divergence, since the endpoints of
the integration in (\ref{b.4}) are the same for all contours. Consequently,
the divergence of $\Delta\mathcal{D}_{j}$ vanishes, $\partial_{j}%
\Delta\mathcal{D}_{j}=0$. It follows that $\Delta\mathcal{D}_{j}$ is the curl
of some vector. Hence, the difference between contours is included in the form
(\ref{b.6}).

\section{Percus pressure tensor}

\label{ap:C}The free energy is obtained from the grand potential by a Legendre
transform%
\begin{equation}
\beta F\left(  \beta,V\mid\overline{n}\right)  =-Q^{e}\left(  \beta,V\mid
\nu\right)  +\int d\mathbf{r}\overline{n}\left(  \mathbf{r}\right)  \nu\left(
\mathbf{r}\right)  \label{c.1}%
\end{equation}
where $\overline{n}\left(  \mathbf{r}\right)  =\left\langle n({\mathbf{r}%
})\right\rangle $ is the average number density. The free energy and grand
potential are expressed as integrals of their respective densities%
\begin{equation}
F\left(  \beta,V\mid\overline{n}\right)  =\int d\mathbf{r}f\left(
\mathbf{r},\beta\mid\overline{n}\right)  ,\hspace{0.2in}Q^{e}\left(
\beta,V\mid\nu\right)  =\int d\mathbf{r}\beta p_{T}\left(  \mathbf{r}%
,\beta\mid\nu\right)  \label{c.2}%
\end{equation}
so that
\begin{equation}
\beta p_{T}\left(  \mathbf{r},\beta\mid\nu\right)  =\overline{n}\left(
\mathbf{r}\right)  \nu\left(  \mathbf{r}\right)  -f\left(  \mathbf{r}%
,\beta\mid\overline{n}\right)  .\label{c.3}%
\end{equation}
The Percus pressure tensor \cite{Percus86} is defined by%
\begin{equation}
p_{ij}\left(  \mathbf{r},\beta\mid\overline{n}\right)  \equiv\delta_{ij}%
p_{T}\left(  \mathbf{r},\beta\mid\nu\right)  +\int d\mathbf{r}^{\prime}%
r_{i}^{\prime}\int_{0}^{1}d\gamma\frac{\delta f\left(  \mathbf{r}%
+\gamma{\mathbf{r}}^{\prime}-{\mathbf{r}}^{\prime},\beta\mid\overline
{n}\right)  }{\delta\overline{n}(\mathbf{r}+\gamma{\mathbf{r}}^{\prime}%
)}\partial_{j}\overline{n}\left(  \mathbf{r}+\gamma\mathbf{r}^{\prime}\right)
.\label{c.4}%
\end{equation}
The corresponding mechanical pressure is
\begin{equation}
p_{m}\left(  \mathbf{r},\beta\mid\overline{n}\right)  =p_{T}\left(
\mathbf{r},\beta\mid\nu\right)  +\frac{1}{3}\int d\mathbf{r}^{\prime}\int
_{0}^{1}d\gamma\frac{\delta f\left(  \mathbf{r}+\gamma{\mathbf{r}}^{\prime
}-{\mathbf{r}}^{\prime},\beta\mid\overline{n}\right)  }{\delta\overline
{n}(\mathbf{r}+\gamma{\mathbf{r}}^{\prime})}r_{i}^{\prime}\partial
_{i}\overline{n}\left(  \mathbf{r}+\gamma\mathbf{r}^{\prime}\right)
.\label{c.5}%
\end{equation}
No motivation nor interpretation for this result is provided.

First, prove the force balance equation. Separate the pressure tensor into two
parts%
\begin{equation}
p_{ij}\left(  \mathbf{r},\beta\mid\overline{n}\right)  =\delta_{ij}%
p_{T}\left(  \mathbf{r},\beta\mid\overline{n}\right)  +\Delta P_{ij}\left(
\mathbf{r},\beta\mid\overline{n}\right)  , \label{c.6}%
\end{equation}%
\[
\Delta P_{ij}\left(  \mathbf{r},\beta\mid\overline{n}\right)  =\int
d\mathbf{r}^{\prime}r_{i}^{\prime}\int_{0}^{1}d\gamma\frac{\delta f\left(
\mathbf{r}+\gamma{\mathbf{r}}^{\prime}-{\mathbf{r}}^{\prime},\beta
\mid\overline{n}\right)  }{\delta\overline{n}(\mathbf{r}+\gamma{\mathbf{r}%
}^{\prime})}\partial_{j}\overline{n}\left(  \mathbf{r}+\gamma\mathbf{r}%
^{\prime}\right).
\]
Then%
\begin{align}
\delta_{ij}\partial_{i}\beta p_{T}\left(  {\mathbf{r}},\beta\mid\overline
{n}\right)   &  =\delta_{ij}\partial_{i}\left(  \overline{n}({\mathbf{r}}%
)\nu({\mathbf{r}})-\beta f\left(  {\mathbf{r}},\beta\mid\overline{n}\right)
\right) \nonumber\\
&  =\overline{n}({\mathbf{r}})\partial_{j}\nu({\mathbf{r}})+\nu({\mathbf{r}%
})\partial_{j}\overline{n}({\mathbf{r}})-\partial_{j}\beta f\left(
\mathbf{r},\beta\mid\overline{n}\right)  \label{c.7}%
\end{align}
and%
\begin{align}
\partial_{i}\beta\Delta p_{ij}\left(  \mathbf{r},\beta\mid\overline{n}\right) 
&  =\int d\mathbf{r}^{\prime}r_{i}^{\prime}\partial_{i}\int_{0}^{1}%
d\gamma\frac{\delta\beta f\left(  \mathbf{r}+\gamma{\mathbf{r}}^{\prime
}-{\mathbf{r}}^{\prime},\beta\mid\overline{n}\right)  }{\delta\overline
{n}(\mathbf{r}+\gamma{\mathbf{r}}^{\prime})}\partial_{j}\overline{n}\left(
\mathbf{r}+\gamma\mathbf{r}^{\prime}\right) 
\nonumber \\
&  =\int d\mathbf{r}^{\prime}\int_{0}^{1}d\gamma\partial_{\gamma}\left(
\frac{\delta\beta f\left(  \mathbf{r}+\gamma{\mathbf{r}}^{\prime}-{\mathbf{r}%
}^{\prime},\beta\mid\overline{n}\right)  }{\delta\overline{n}(\mathbf{r}%
+\gamma{\mathbf{r}}^{\prime})}\partial_{j}\overline{n}\left(  \mathbf{r}%
+\gamma\mathbf{r}^{\prime}\right)  \right) 
\nonumber \\
&  =\int d\mathbf{r}^{\prime}\left(  \frac{\delta\beta f\left(  \mathbf{r}%
,\beta\mid\overline{n}\right)  }{\delta\overline{n}(\mathbf{r}+{\mathbf{r}%
}^{\prime})}\partial_{j}\overline{n}\left(  \mathbf{r}+\mathbf{r}^{\prime
}\right)  -\frac{\delta\beta f\left(  \mathbf{r}-{\mathbf{r}}^{\prime}%
,\beta\mid\overline{n}\right)  }{\delta\overline{n}(\mathbf{r})}\partial
_{j}\overline{n}\left(  \mathbf{r}\right)  \right)
\nonumber \\
&  =\int d\mathbf{r}_{1}\frac{\delta\beta f\left(  \mathbf{r},\beta
\mid\overline{n}\right)  }{\delta\overline{n}(\mathbf{r}_{1})}\partial
_{1j}\overline{n}\left(  \mathbf{r}_{1}\right)  -\frac{\delta\beta F\left(
\beta\mid\overline{n}\right)  }{\delta\overline{n}(\mathbf{r})}\partial
_{j}\overline{n}\left(  \mathbf{r}\right) \nonumber
\nonumber \\
&  =-\left(  \nu({\mathbf{r}})\partial_{j}\overline{n}\left(  \mathbf{r}%
\right)  -\partial_{j}\beta f\left(  \mathbf{r},\beta\mid\overline{n}\right)
\right)   \label{c.8}%
\end{align}
so the force balance equation is verified%
\begin{equation}
\partial_{i}\beta p_{ij}\left(  \mathbf{r},\beta\mid\overline{n}\right)
=\overline{n}({\mathbf{r}})\partial_{j}\nu({\mathbf{r}}). \label{c.9}%
\end{equation}

For the pressure to qualify as a thermodynamic pressure it should satisfy%
\begin{equation}
\int d\mathbf{r}\beta p_{m}\left(  \mathbf{r},\beta\mid\nu\right)
=Q^{e}\left(  \beta,V\mid\nu\right)  .\label{c.10}%
\end{equation}
Use of the Percus form (\ref{c.5}) gives this condition to be%
\begin{equation}
\int d\mathbf{r}\frac{1}{3}\int d\mathbf{r}^{\prime}\int_{0}^{1}d\gamma
\frac{\delta f\left(  \mathbf{r}+\gamma{\mathbf{r}}^{\prime}-{\mathbf{r}%
}^{\prime},\beta\mid\overline{n}\right)  }{\delta\overline{n}(\mathbf{r}%
+\gamma{\mathbf{r}}^{\prime})}r_{i}^{\prime}\partial_{i}\overline{n}\left(
\mathbf{r}+\gamma\mathbf{r}^{\prime}\right)  =0.\label{c.11}%
\end{equation}
The Appendix of Ref. \citenum{Pozhar} claims that this is true, but no proof
is given.